\documentclass[aps,prb,preprint,superscriptaddress]{revtex4-2}
\usepackage{amsfonts}
\usepackage{amssymb}
\usepackage{amsmath}
\usepackage{graphicx}
\usepackage{dcolumn}
\usepackage{bm}
\usepackage{units}
\usepackage{multirow}
\usepackage{CJKutf8}
\usepackage{subfigure}
\usepackage{color}%
\usepackage{url}
\usepackage[colorlinks,linkcolor=red,anchorcolor=green,citecolor=blue]{hyperref}
\providecommand{\U}[1]{\protect\rule{.1in}{.1in}}
\usepackage{array}
\newcommand{\PreserveBackslash}[1]{\let\temp=\\#1\let\\=\temp}
\newcolumntype{C}[1]{>{\PreserveBackslash\centering}p{#1}}
\newcolumntype{R}[1]{>{\PreserveBackslash\raggedleft}p{#1}}
\newcolumntype{L}[1]{>{\PreserveBackslash\raggedright}p{#1}}
\allowdisplaybreaks

\begin{document}


\title{On the angular dependence of anomalous Hall current}


\author{Lulu Li}
\affiliation{Key Laboratory of Quantum Materials and Devices of Ministry of Education, School of Physics, Southeast University, Nanjing 211189, China}

\author{Junwen Sun}
\affiliation{Physics Department, The Hong Kong University of Science and Technology, Clear Water Bay, Kowloon, Hong Kong}
\affiliation{HKUST Shenzhen Research Institute, Shenzhen 518057, China}

\author{Lei Wang}
\email{wanglei.icer@seu.edu.cn}
\affiliation{Key Laboratory of Quantum Materials and Devices of Ministry of Education, School of Physics, Southeast University, Nanjing 211189, China}

\author{X. R. Wang}
\affiliation{Physics Department, The Hong Kong University of Science and Technology, Clear Water Bay, Kowloon, Hong Kong}
\affiliation{School of Science and Engineering, Chinese University of Hong Kong, Shenzhen, Shenzhen 518172, China}

\author{Ke Xia}
\email{kexia@seu.edu.cn}
\affiliation{Key Laboratory of Quantum Materials and Devices of Ministry of Education, School of Physics, Southeast University, Nanjing 211189, China}


\begin{abstract}
The transverse current ($\mathbf{j}_H$) due to anomalous Hall effect (AHE) is usually assumed to be perpendicular to the magnetization ($\mathbf{m}$) in ferromagnetic materials, which governs the experiments in spintronics. Generally, this assumption is derived from a continuum model, where the crystal's discrete symmetry is effectively represented by the concept of an effective mass from the band structure. In this paper, we calculate the spin transport through the nonmagnetic metal (NM) $\vert$ ferromagnetic metal (FM) interfaces and find that the corresponding Hall current is generally not perpendicular to $\mathbf{m}$ with only a few exceptions at high symmetry crystal orientations. The calculation illustrates the breakdown of $\mathbf{j}_H=\Theta \mathbf{m}\times\mathbf{j}_c$, where $\Theta$ denotes the anomalous Hall angle and $\mathbf{j}_c$ represents the injecting charge current. An analytical formula based on the discrete symmetry of the solid can describe this effect well. In this framework, the leading order corresponds to the conventional AHE, while higher-order terms account for deviations in the Hall current. Additionally, we identify the presence of a chiral anomalous Hall effect (CAHE) at interface with odd rotational symmetry (e.g., $C_{3v}$) and the higher-order terms can even dominate the AHE by constructing superlattices. The general existence of hidden chirality in spin transport is also revealed, with a specific focus on interface chirality (IC). Our results highlight the significance of discrete atomic positions in solids for spin transport, which extends beyond the conventional continuum model. Moreover, considering the important application of the AHE in spintronics and the wide existence of the interfaces in the devices, the breakdown of $\mathbf{j}_H=\Theta \mathbf{m}\times\mathbf{j}_c$ suggests that all experimental measurements related to the AHE should be re-evaluated.
\end{abstract}


\maketitle

\section*{Introduction}
As an efficient method to measure the magnetism, the anomalous Hall effect~\cite{a2,a3,a6,a7,a4,RevModPhys.82.1959} (AHE) is crucial in spintronics. Conventionally, the corresponding measurements in devices related to AHE is based on the product rule, reads,
\begin{eqnarray}
	\mathbf{j}_H=\Theta \mathbf{m}\times\mathbf{j}_c
\end{eqnarray}
where $\mathbf{j}_H$ represents the measured anomalous Hall current, $\Theta$ is the anomalous Hall angle of the material, $\mathbf{m}$ stands for the direction of magnetization, and $\mathbf{j}_c$ is the injecting charge current. Generally, this product rule governs the experimental measurements of magnetism and the corresponding studies of AHE mainly focus on the magnitude of $\Theta$ for emerging materials~\cite{Zhou2022,Wang2023,Krempask2024,doi:10.1126/science.adf1506,Ikhlas2022,Tseng2022,Takagi2023,Pan2022,Tang2022,Feng2022,doi:10.1126/sciadv.adj4883,PhysRevLett.130.126302,PhysRevLett.130.166702,PhysRevLett.130.036702,PhysRevLett.132.026301} and their original mechanisms~\cite{a4,RevModPhys.82.1959}, such as the intrinsic Berry curvature~\cite{Fang92,PhysRevLett.92.037204,Yoo2021,Wang2018,PhysRevMaterials.8.084201,PhysRevLett.129.185301}, skew scattering~\cite{SMIT1955877,SMIT195839,doi:10.1126/sciadv.abb6003,PhysRevLett.131.076601,Fujishiro2021}, side jump~\cite{PhysRevB.2.4559,PhysRevB.83.125122,PhysRevB.99.245418,Li_2015}, and their corresponding scaling law~~\cite{PhysRevLett.103.087206,PhysRevLett.114.217203,wang2021first,wang2023abnormal,Siddiquee2023}. 

The above assumption is based on a continuum model, in which the discrete symmetry of the crystal is effectively captured by the mass derived from the band structure. However, interfaces are ubiquitous in spintronic devices and play a critical role in various spin transport phenomena, which extend beyond the scope of the continuum model with effective mass. These include the giant interface spin Hall and anomalous Hall effects~\cite{PhysRevLett.116.196602,PhysRevLett.121.136805,PhysRevB.109.104422,Dai2024,PhysRevB.108.L241403}, spin injection through interfaces for detecting spin currents~\cite{Hautzinger2024,Min2022,PhysRevLett.97.026602,Guarochico-Moreira2022,PhysRevApplied.15.054018}, spin Hall magnetoresistance~\cite{PhysRevLett.110.206601,PhysRevLett.116.097201} for sensitive magnetic field sensors~\cite{admt.201800073}, the spin filter effect for giant tunneling magnetoresistance~\cite{doi:10.1126/science.aar4851}, and spin-orbit torque for efficient magnetic switching~\cite{Yang2024,Xue2023,doi:10.1126/sciadv.aax8467,adma.201705699,Zheng2021,RevModPhys.91.035004}, among others. Additionally, high-quality interfaces are crucial for optimizing the performance of spintronic devices, and recent experiments have achieved such high-quality interfaces~\cite{PhysRevX.14.021045}, making the study of spin transport through interfaces increasingly urgent. Therefore, it is important to verify the applicability of the AHE product rule in metallic interfaces to address any potential gaps in conventional measurements.

In this paper, we investigate an interface between nonmagnetic metal (NM) and ferromagnetic metal (FM) as an efficient system and use first-principle calculations to study the anomalous Hall effect (AHE) in the presence of interface chirality (IC). Our results show that the conventional continuum model for the AHE is insufficient, as the Hall current deviates from the usual product rule in both direction ($\mathbf{j}_H\nparallel\mathbf{m}\times\mathbf{j}_c$) and magnitude ($\mathbf{j}_H(\mathbf{m})\neq-\mathbf{j}_H(-\mathbf{m})$). We develop an analytical theory to describe such effect based on the discrete symmetry of the solid, where the leading order corresponds to the conventional AHE, and higher-order terms related to rotational symmetry account for deviations in the Hall current, consistent with the theory of the multipolar structure of the Berry curvature~\cite{peng2024observationinplaneanomaloushall,liu2024multipolaranisotropyanomaloushall}. Furthermore, we find that the odd rotational symmetry ($C_{3v}$) results in a change in the magnitude of the Hall current upon switching the magnetization, corresponding to the chiral anomalous Hall effect (CAHE) and the higher-order terms can even dominate the AHE rather than the conventional AHE by constructing superlattices. Additionally, we also reveal that this hidden chirality is particularly prominent in spin transport within the lateral Brillouin zone (BZ). When crystal symmetry is taken into account, the total conductance through interfaces remains equivalent for both left- and right-hand cases, resulting in hidden chirality in longitudinal conductance measurements. Due to the differing symmetries between longitudinal and transverse transports, the transverse AHE can reveal this chirality effectively. These findings highlight the importance of discrete atomic positions in solids and their associated crystal symmetry in spin transport.

\section*{Results}

\begin{figure}[tp]
	\includegraphics[width=0.8\columnwidth]{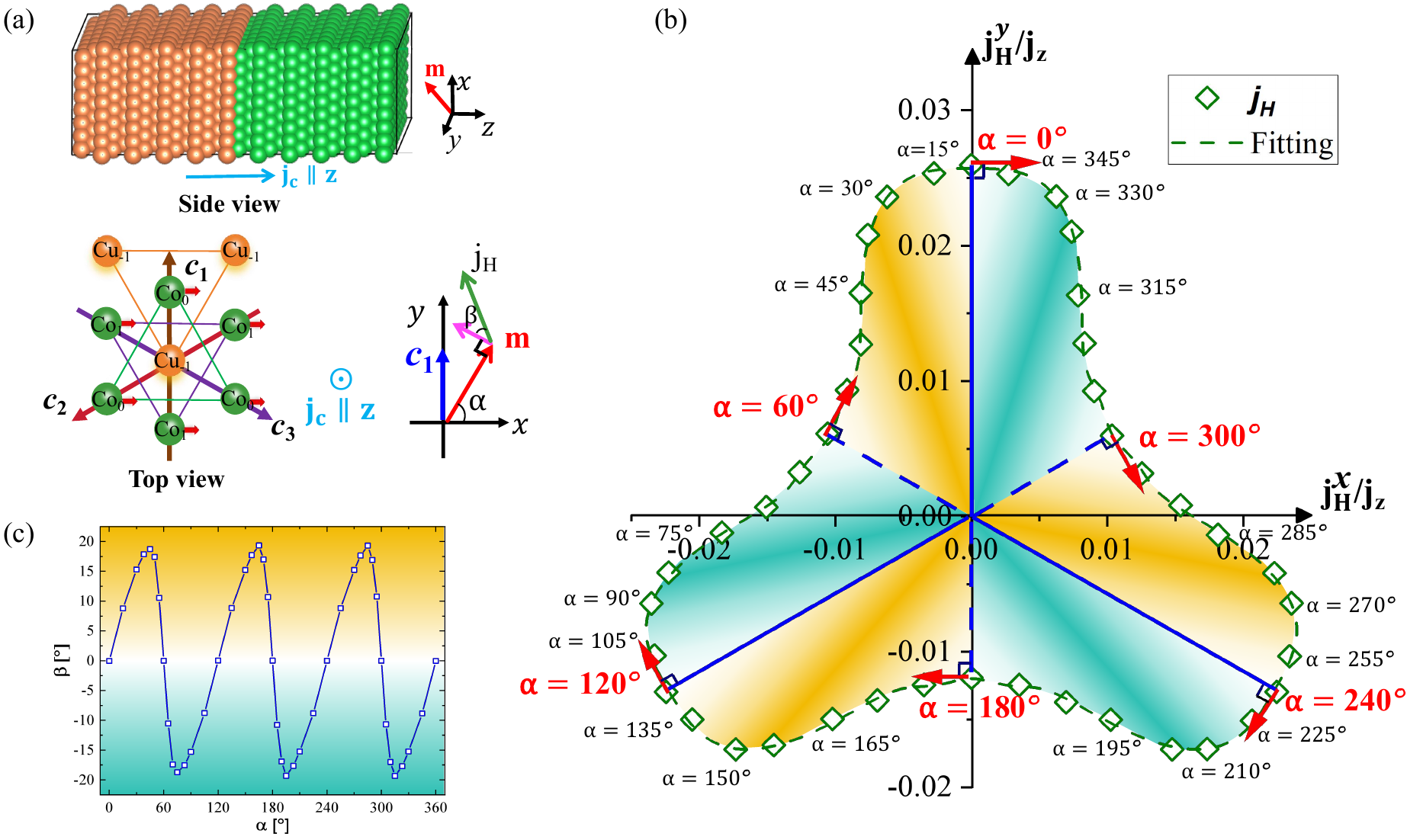}
	\caption{(a) The side and top views of the Cu$\vert$Co interfaces along (111) direction. The charge current is injected along the fcc (111) direction ($z$-axis). The magnetization ($\mathbf{m}$) is rotated by an angle $\alpha$ in the ${xy}$-plane, and $\mathbf{c}_{1,2,3}$ represent the translation vectors of the fcc (111) plane. The corresponding Hall current $\mathbf{j}_H$ is also indicated by a green arrow, forming an angle $\beta$ with $\mathbf{m}\times\mathbf{j}_c$ (pink arrow). (b) The angle ($\alpha$) dependence of the anomalous Hall currents ($j_H^y$ and $j_H^x$) for the first Co layer around the Cu$\vert$Co interface along fcc (111) direction. The red arrows indicate the direction of the magnetization $\mathbf{m}$. The dashed line represents the fitting results obtained using Eq.~(\ref{fit}) with $n=3$. (c) The $\alpha$ versus the deviation angle ($\beta$) between the directions of the calculated anomalous Hall current $\mathbf{j}_H^i=(j_H^x,j_H^y)$ and the conventional $\mathbf{m}\times\mathbf{j}_c$.}
	\label{angle}
\end{figure}

A systematic study of the anomalous Hall current ($\mathbf{j}_H$) at Cu$\vert$Co interfaces is conducted to examine the product rule of the AHE. In this study, the injected charge current $\mathbf{j}_c$ is aligned along the $z$-axis and fixed in the fcc (111) direction, while the magnetization vector $\mathbf{m}$ is rotated in the $xy$-plane by an angle $\alpha$, as illustrated in Fig.~\ref{angle} (a). The corresponding transport properties are calculated from the homemade first principles method using the fully relativistic exact muffin-tin orbitals (FR-EMTO)~\cite{wang2021first,wang2022crystal,wang2023abnormal} within the framework of scattering wave functions~\cite{PhysRevB.44.8017,PhysRevB.73.064420,PhysRevB.97.214415,PhysRevB.99.144409}. It should be noted that $\mathbf{j}_H$ can be evaluated for any specific atomic layer inside the system. In this paper, unless otherwise specified, we present the calculated results for the first Co layer near the interface, labeled as $\mathbf{j}_H^i$.

As illustrated in the lower panel of Fig.~\ref{angle} (a), there are three translation vectors for the fcc (111) plane that can be used to construct the fcc crystal structure, denoted as $\mathbf{c}_{1,2,3}$, which give rise to a triple rotational symmetry ($C_{3v}$) about the $z$-axis. In Fig.~\ref{angle} (b), we plot the components of the anomalous Hall current, $\mathbf{j}_H^i=(j_H^x,j_H^y)$, in the first Co layer near the interface for various orientations of $\mathbf{m}$ within the $xy$-plane. It is evident that as $\mathbf{m}$ rotates in the $xy$-plane ($\alpha\in[0,360]$), the components of the anomalous Hall current, $\mathbf{j}_H^i=(j_H^x,j_H^y)$, conform to the $C_{3v}$ symmetry derived from $\mathbf{c}_{1,2,3}$.

Furthermore, we notice that, except when $\mathbf{j}_H^i$ is parallel to the projections of $\mathbf{c}_{1,2,3}$ onto the $xy$-plane, the direction of the anomalous Hall current is not $\mathbf{m}\times\mathbf{j}_c$. To illustrate this effect more clearly, we define a deviation angle ($\beta$) as depicted in the Fig.~\ref{angle} (a), which describes the angle between the direction of $\mathbf{j}_H^i=(j_H^x,j_H^y)$ and $\mathbf{m}\times\mathbf{j}_c$, thus $\beta=0$ signifies for a conventional relation of $\mathbf{j}_H=\Theta\mathbf{m}\times\mathbf{j}_c$. The corresponding results are plotted in Fig.~\ref{angle} (c), where $\beta$ oscillates in the range of $\beta \in [-20,20]$ and exhibits triple rotational symmetry. This deviation arises from the differing magnitudes of anomalous Hall currents during magnetization switching ($\mathbf{j}_H(\mathbf{m})\neq-\mathbf{j}_H(-\mathbf{m})$), as shown by the long solid lines and short dashed lines, where the primary axes ($\pm\mathbf{c}_{1,2,3}$) for a given $\mathbf{j}_H^i$ are inequivalent. Thus, when the direction of the anomalous Hall current deviates from the primary axes ($\pm\mathbf{c}_{1,2,3}$) of the fcc (111) lattice, the crystal induce transverse current~\cite{wang2022crystal} appears and results in the nonzero $\beta$. Introducing random disorders equalizes the differences between these primary axes, resulting in the direction of the anomalous Hall current eventually adhering to $\mathbf{j}_H=\Theta\mathbf{m}\times\mathbf{j}_c$~\cite{si}, consistent with the above hypothesis.

Given the significance of the crystal structure of the fcc (111) plane in the interface induced AHE as discussed above, we employ the tensor theory~\cite{rug01:000711810,1058782,WangXR2023,10.1063/5.0187589} to describe this deviated Hall current, where, the linear response between the longitudinal and transverse charge current can be expressed as $\mathbf{j}_H=\overleftrightarrow{\rho}(\mathbf{m})\mathbf{j}_c$. The linear response coefficients $\overleftrightarrow{\rho}(\mathbf{m})$ can be expanded in terms of $\mathbf{m}$ as follows:
\begin{eqnarray}
	\overleftrightarrow{\rho}_{ij}(\mathbf{m})=\rho_{ij}^0+\rho_{ijk}m_k+\rho_{ijkl}m_km_l+\cdots
\end{eqnarray}
where $\rho_{ijk}=\partial\rho_{ij}/\partial m_k$, $i,j,k\in\{x,y,z\}$, and an Einstein summation convention is used. In our system, $\mathbf{j}_c=(0,0,j_z)$ and $\mathbf{m}=(\cos\alpha,\sin\alpha,0)$. As illustrated in the top view of the interface in Fig.~\ref{angle} (a), the in-plane point group of the crystal structure exhibits $C_{3v}$ symmetry related to $\mathbf{c}_{1,2,3}$ and mirror symmetry ($\mathcal{M}_y$) with respect to the $y$-axis. This leads to constraints on $\overleftrightarrow{\rho}_{ij}$, allowing us to ultimately derive~\cite{si},
\begin{eqnarray}
	\mathbf{j}_H/j_z=\left[\begin{array}{c}
		j_H^x/j_z \\
		j_H^y/j_z
	\end{array}\right]=\sum_{i=3n-2}^{n\in[1,\infty)}p_{i}\left[\begin{array}{c}
		\sin i\alpha \\
		-\cos i\alpha 
	\end{array}\right]+\sum_{j=3n-1}^{n\in[1,\infty)}p_{j}\left[\begin{array}{c}
		\sin j\alpha \\
		\cos j\alpha 
	\end{array}\right].
	\label{fit}
\end{eqnarray}
Here, we omit the $z$ component of $\mathbf{j}_H$ due to the Hall current always flows inside the $xy$-plane and $p_{i/j}$ are the combination of different $\rho_{ijk}$, $\rho_{ijkl}$, and so on~\cite{si}. Clearly, the first term with $i=1$ corresponds to the conventional AHE, given by the product rule $\mathbf{j}_H=p_1 \mathbf{m}\times\mathbf{j}_c$. In this sense, remaining terms with $i\neq1$ contribute to the deviation of $\mathbf{j}_H$. By fitting our calculated $\mathbf{j}_H$ in Fig.~\ref{angle} (b) using Eq.~(\ref{fit}), we determine that $n=3$ provides a sufficiently accurate representation, as indicated by the corresponding dashed line. The fitting parameters are listed in Tab.~\ref{pars} for comparison. Moreover, the perpendicular condition between $\mathbf{j}_H$ and $\mathbf{m}$ can be expressed as $\mathbf{j}_H\cdot\mathbf{m}\propto\sin3n\alpha_p=0$. This indicates that $\alpha_p=l\pi/3$ for integer values of $l$, resulting in six configurations where the two vectors are perpendicular, even with nonzero values of $p_{i,i\neq1}$. A similar deviation of the Hall current can also be observed at the interfaces formed by artificial hcp crystal structures~\cite{si}. Therefore, the in-plane symmetry and the local vectors of these interfaces are crucial for this effect on a microscopic level.

\begin{table}
	\begin{tabular}{ccccc}
		\hline\hline 
		Interfaces & $p_1$ & $p_2$ & $p_4$ & $p_5$  \\ \hline
		Cu$\vert$Co fcc (111) & -0.021 & 0.0065 & - & -0.0017  \\
		Cu$\vert$Co hcp (0001) & 0.095 & -0.0028 & - & -   \\ 
		\textbf{[}Cu$_2$$\vert$Co$_2$\textbf{]}$_n$ fcc (111) & -0.0077 & 0.061 & -0.0025 & - \\ \hline
		& $p_1$ & $p_3$ & $p_5$ & $p_7$  \\ \hline
		Cu$\vert$Co fcc (001) & -0.019 & -& -& -\\
		\hline \hline
	\end{tabular}
	\caption{The fitting parameters for different interfaces are presented. We omit the small values ($p_i<1\times10^{-3}$) to facilitate a clearer comparison between them.}
	\label{pars}
\end{table}

We also observe that $\mathbf{j}_H(\mathbf{m})\neq-\mathbf{j}_H(-\mathbf{m})$, indicating the presence of a CAHE, which may be attributed to the interface chirality (IC). As is well known, the presence of a symmetry-breaking electric field $\mathbf{E}$ at the interface inevitably breaks the inversion symmetry. This induces interface spin-orbit coupling described by $H_{\mathbf{soc}}=\lambda_{soc}\hat{\sigma}\cdot\left(\mathbf{E}\times\mathbf{k}\right)$, where $\lambda_{soc}$ represents the spin-orbit coupling strength, $\hat{\sigma}$ denotes the Pauli matrices vector, and $\mathbf{k}$ is the wave vector. This effect generally leads to significant chiral interface states, which are crucial for topological quantized transport~\cite{Zhang2024,Zhao2023,PhysRevB.94.165443,PhysRevA.102.023520,PhysRevB.92.075138}. Given that the AHE is sensitive to magnetization, which is vital for chirality in spintronics~\cite{Yang2021,cheong2022magnetic2,yu2023chirality,yu2021magnetic4,gobel2021beyond5}, a CAHE should be present when considering the IC and results in the above deviation of spin Hall currents. In fact, chirality should naturally arise at interface for any spin transport, even in the absence of quantized channels. We will further illustrate the universality of this chiral spin transport with IC and explain why such chirality is obscured in experiments for specific crystal orientations through symmetry arguments.

\begin{figure*}[tp]
	\includegraphics[width=0.8\columnwidth]{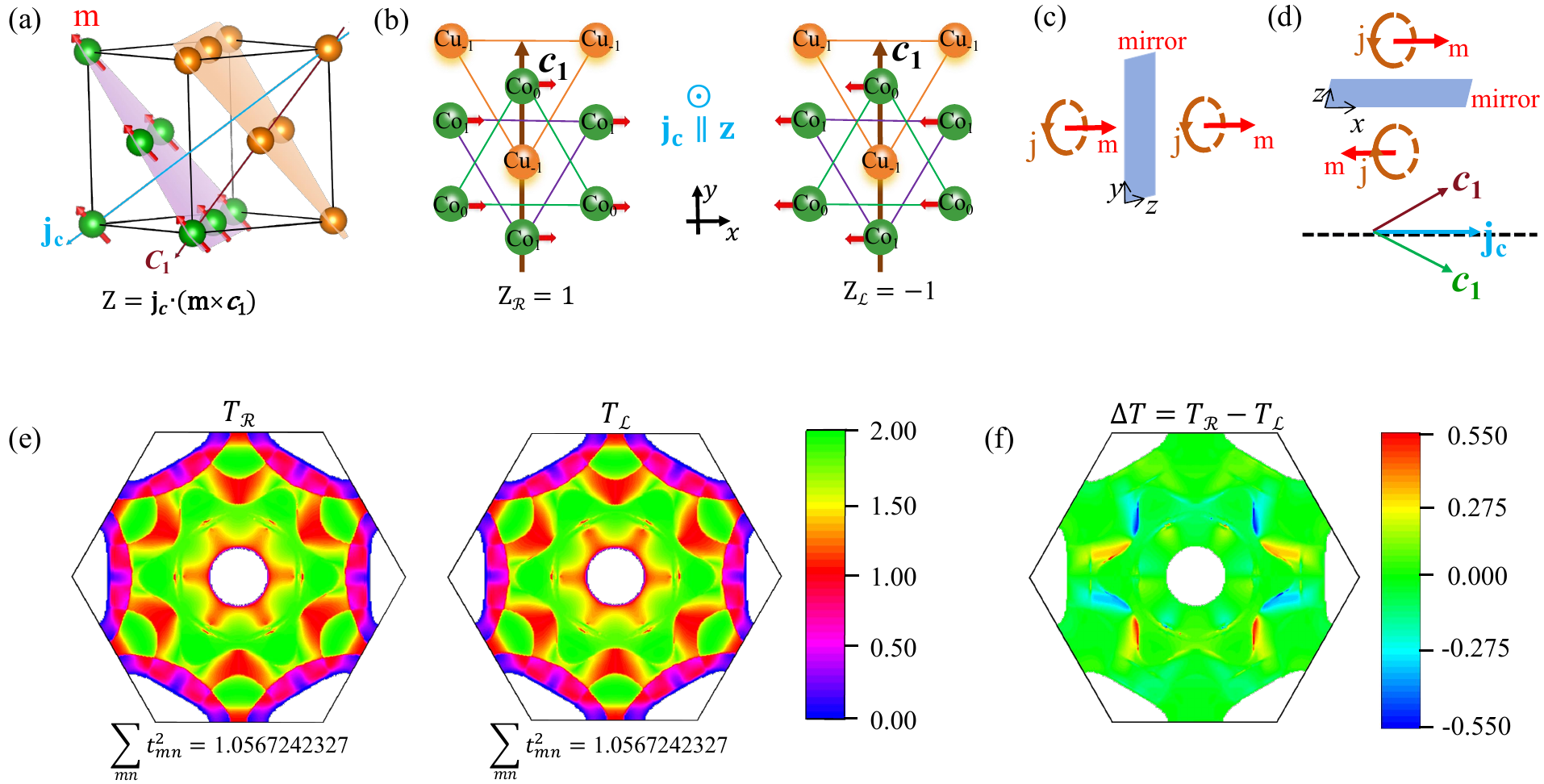}
	\caption{(a) The fcc structure features a Cu$\vert$Co interface oriented along the (111) direction, defining the chirality as $\rm{Z}$=$\mathbf{j}_c$$\cdot(\mathbf{m}\times\mathbf{c}_1)$, where $\mathbf{j}_c$ represents the longitudinal charge current, $\mathbf{m}$ denotes the direction of the magnetization, and $\mathbf{c}_1$ signifies the crystal direction of the three inequivalent planes along the fcc (111) orientation. (b) shows the top view of the interfaces for right-hand ($\rm{Z}_{\mathcal{R}}=1$) and left-hand ($\rm{Z}_{\mathcal{L}}=-1$) cases, respectively. (c) and (d) illustrate the mirror symmetry analysis.  (e) displays the $\mathbf{k}_{\parallel}$-resolved transmission probabilities ($T_{\mathcal{R/L}}$) in the lateral BZ. (f) The difference of transmission probabilities $\Delta T$ = $T_{\mathcal{R}}$ - $T_{\mathcal{L}}$.}
	\label{trans}
\end{figure*}

As depicted in Fig.~\ref{trans} (a), for the Cu$\vert$Co interface along (111) direction, the direction of magnetization ($\mathbf{m}$), one of the translation vectors of the (111) plane ($\mathbf{c}_1$), and the charge current ($\mathbf{j}_{c}$) are utilized to construct the vector products of IC with its chiral index defined as $\rm{Z}$=$\mathbf{j}_{c}$$\cdot (\mathbf m \times \mathbf{c}_{1})$. When switching the direction of the magnetization (e.g. $\mathbf{m}\parallel x$ or $\mathbf{m}\parallel -x$), the distinct right- or left- hand states with $\rm{Z}_{\mathcal{R/L}}=\pm1$ are observed, as depicted in Fig.~\ref{trans} (b), respectively.

The calculated transmission probabilities $T_{\mathcal{R}}(\mathbf{k}_{\parallel})$ and $T_{\mathcal{L}}(\mathbf{k}_{\parallel})$ of the Cu$\vert$Co interface for $\rm{Z}_{\mathcal{R}}=1$ and $\rm{Z}_{\mathcal{L}}=-1$ are displayed in Fig.~\ref{trans} (e), respectively, with $\mathbf{k}_{\parallel}$ representing the wave vector in the lateral BZ. Moreover, for a clearer comparison between $T_{\mathcal{R}}(\mathbf{k}_{\parallel})$ and $T_{\mathcal{L}}(\mathbf{k}_{\parallel})$, $\Delta T(\mathbf{k}_{\parallel})=T_{\mathcal{R}}(\mathbf{k}_{\parallel})-T_{\mathcal{L}}(\mathbf{k}_{\parallel})$ is also plotted in Fig.~\ref{trans} (f). From these results, the following conclusions can be drawn: 1) the nonzero $\Delta T(\mathbf{k}_{\parallel})$ in Fig.~\ref{trans} (f) indicates the chirality of the $\mathbf{k}_{\parallel}$-resolved transmission probabilities and the largest values of $\Delta T(\mathbf{k}_{\parallel})/T_{\mathcal{L}}(\mathbf{k}_{\parallel})\simeq 78.4\%$ highlight a pronounced chirality at the interface; 2) the total conductances, obtained by integrating $T(\mathbf{k}_{\parallel})$ across the lateral BZ, are found to be identical for $\rm{Z}_{\mathcal{R}}=1$ and $\rm{Z}_{\mathcal{L}}=-1$.

To delve into the concealed chirality of total conductance, we look towards the symmetry of the BZ. As illustrated in Fig.~\ref{trans} (e) and (f), the transmission $T(\mathbf{k}_{\parallel})$ adheres to the symmetry of $T(m_x, k_{x})=T(m_x, -k_{x})$ and $T(m_x, k_{y})=T(-m_x, -k_{y})$, respectively. The first symmetry is characterized by the mirror symmetry inherent to the system itself. As depicted in Fig.~\ref{trans} (b), the in-plane crystal structure clearly exhibits mirror symmetry with respect to the $yz$-plane; when $\mathbf{m}\parallel x$, the magnetization remains unchanged under mirror operations according to the $yz$-plane, as indicated by the molecular current in Fig.~\ref{trans} (c). Thus, the mirror symmetry with respect to the $yz$-plane implies $T(m_x, k_{x})=T(m_x, -k_{x})$. Conversely, when considering the $xz$-plane, the magnetization will be switched by the mirror operation when $\mathbf{m}\parallel x$ as depicted in Fig.~\ref{trans} (d), which gives the symmetry of $T(m_x, k_{y})=T(-m_x, -k_{y})$. Here one may notice that the $\mathbf{c}_1$ in Fig.~\ref{trans} (b) does not exhibit mirror symmetry with respect to the $xz$-plane, however, as illustrated in Fig.~\ref{trans} (d), the projections of $\mathbf{c}_1$ and its mirror image onto $\mathbf{j}_c$ are equivalent, which recover the corresponding mirror symmetry of $\mathbf{j}_c$ and consequently, that of $T(\mathbf{k}_{\parallel})$.

With the previous symmetries and upon summing over the lateral BZ, it is found that $T(\mathbf m) = T(-\mathbf m)$ in any case~\cite{si}. Consequently, despite the $T(\mathbf{k}_{\parallel})$ being chiral in lateral BZ, the chirality of the measurable conductance vanishes in experimental measurements. The quantum well states are well known to be significantly influenced by the $\mathbf{k}_{\parallel}$-resolved reflectivity ($R=1-T$) of the interfaces~\cite{wu2009effect}, the pronounced chirality observed in $T(\mathbf{k}_{\parallel})$  at specific $\mathbf{k}_{\parallel}$ values within the hot spots, as illustrated in Fig.~\ref{trans} (f), can be effectively quantified using the corresponding quantum well state technique by switching the magnetization. 

\begin{figure*}[tp]
	\includegraphics[width=0.8\columnwidth]{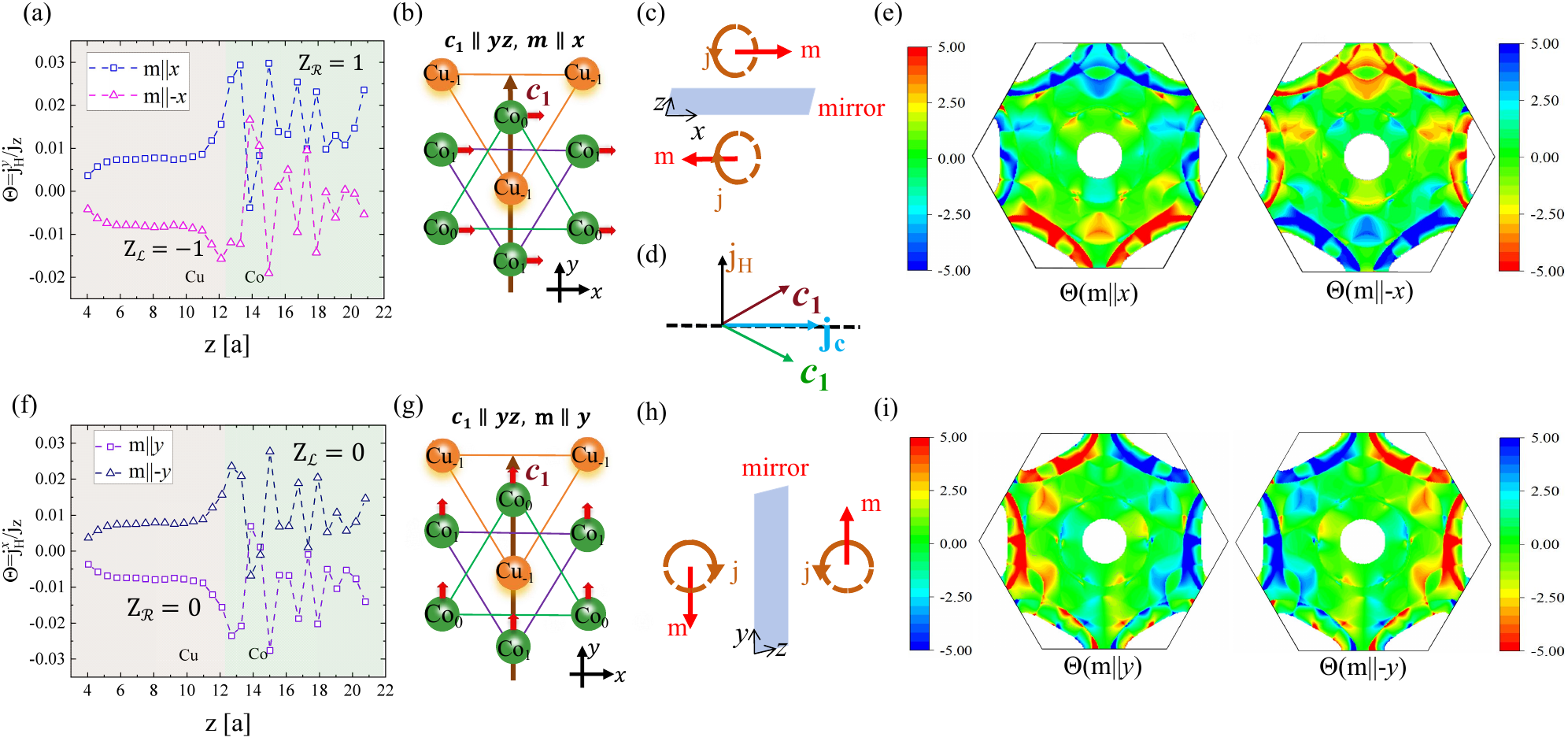}
	\caption{(a) The layer-resolved anomalous Hall angles (${\Theta}^i=\mathbf{j}_H^i/\mathbf{j}_c$ with $i$ representing the layer index) at the Cu$\vert$Co interface along (111) direction, where  ${\mathbf{c}_{1}\parallel yz}$-plane, $\mathbf{j}_c\equiv (0,0,j_z)$, $\mathbf{j}_H^i\equiv (0,j_H^y,0)$ and $\mathbf{m}\parallel \pm x$. (f) The similar anomalous Hall angle with $\mathbf{j}_H^i\equiv (j_H^x,0,0)$ and $\mathbf{m}\parallel \pm y$. (b) and (g) depict the top view of the Cu$\vert$Co interface with $\mathbf{m}\parallel x$ and $\mathbf{m}\parallel y$, respectively. (c), (d) and (h) provide the corresponding sketches for the mirror symmetry analysis of the AHE. (e) and (i) display the $\mathbf{k}_{\parallel}$-resolved $\Theta$ of the first Co layer around interface. Here the color bars are reversed when switching the magnetization to gives a clear view of the mirror symmetry.}
	\label{cahe}
\end{figure*}

Instead of the longitudinal conductance, the total transverse AHE~\cite{a2,a3,a4,a5,a6,a7,a8,a9,PhysRevB.99.224416} should be chiral due to the different symmetry comparing to the longitudinal transport. Similarly to the transmission calculations shown in Fig.~\ref{trans}, our initial focus is on the anomalous Hall angle within the systems where $\rm{Z}_{\mathcal{R}}=1$ and $\rm{Z}_{\mathcal{L}}=-1$. The corresponding layer-resolved anomalous Hall angles (${\Theta}^i=\mathbf{j}_H^i/\mathbf{j}_c$ with $i$ representing the layer index) of the Cu$\vert$Co interface can be obtained, and the ${\Theta}^i$ are depicted in Fig.~\ref{cahe} (a), where $\mathbf{j}_c\equiv(0,0, j_z)$, $\mathbf{j}_H^i\equiv (0,j_H^y,0)$ and $\mathbf{m}\parallel\pm x$, respectively. Based on our findings, when magnetization is switched, the anomalous Hall angles do not exhibit equal and opposite values for $\rm{Z}_{\mathcal{R}}=1$ and $\rm{Z}_{\mathcal{L}}=-1$, indicating discernible and measurable chirality.

To reveal the physical origin of the CAHE, where $\Theta(\mathbf{m})\neq-\Theta(\mathbf{-m})$, similar symmetry arguments are carried out in the following. As we all know, $\mathbf{j}_H\propto \mathbf{m}\times\mathbf{j}_c$ conventionally, underscoring that mirror symmetry with respect to the $\mathbf{m}$-$\mathbf{j}_c$ plane is crucial for the chirality of the AHE. Specifically, for the system depicted in Fig.~\ref{cahe} (b), the $\mathbf{m}$-$\mathbf{j}_c$ plane aligns with the $xz$-plane. Building on the previous analysis of the mirror symmetry, as depicted in Fig.~\ref{cahe} (b) and (c), the magnetization will be reversed by the mirror operation with respect to the $xz$-plane, consequently, one would anticipate a symmetry of $\Theta(m_x,k_{x})=-\Theta(-m_x,-k_{x})$, which ensure the conventional law of AHE in ferromagnetic materials: $\Theta(\mathbf{m})=-\Theta(\mathbf{-m})$~\cite{si}. However, as shown in Fig.~\ref{cahe} (b), the $\mathbf{c}_1$ breaks the mirror symmetry of the crystal structure according to $xz$-plane; as illustrated in Fig.~\ref{cahe} (d), the projections of $\mathbf{c}_1$ and its mirror onto $\mathbf{j}_H$ are inequivalent. Therefore, the mirror symmetry of the AHE cannot be restored, resulting in $\Theta(m_x,k_{x})\neq-\Theta(-m_x,-k_{x})$, which induces the chirality of AHE. Fig.~\ref{cahe} (e) plots the $\Theta(\mathbf{k}_{\parallel})$ in the lateral BZ and confirms the broken of the mirror symmetry accordingly.

To corroborate the preceding discussion, we switch the magnetization to $\pm y$ direction, in which $\rm{Z}_{\mathcal{R}}=\rm{Z}_{\mathcal{L}}=0$. The calculated anomalous Hall angles are depicted  in Fig.~\ref{cahe} (f) with $\mathbf{j}_c\equiv (0,0,j_z)$ and $\mathbf{j}_H^i\equiv (j_H^x,0,0)$. It can be observed that the anomalous Hall angles are equal and opposite upon reversing the magnetization, thereby nullifying the chirality of AHE. This can be also understood by the mirror symmetry with respect to the $yz$-plane as shown in Fig.~\ref{cahe} (g) and (h),  where the crystal structure adheres to mirror symmetry and magnetization reverses accordingly under the mirror operation. In the sense, we have $\Theta(m_y,k_{x})=-\Theta(-m_y,-k_{x})$ and finally $\Theta(m_y)=-\Theta(-m_y)$~\cite{si}. Fig.~\ref{cahe} (i) displays the $\Theta(\mathbf{k}_{\parallel})$ in the lateral BZ and reinforces the aforementioned symmetry arguments.

\begin{figure}[tp]
	\includegraphics[width=0.8\columnwidth]{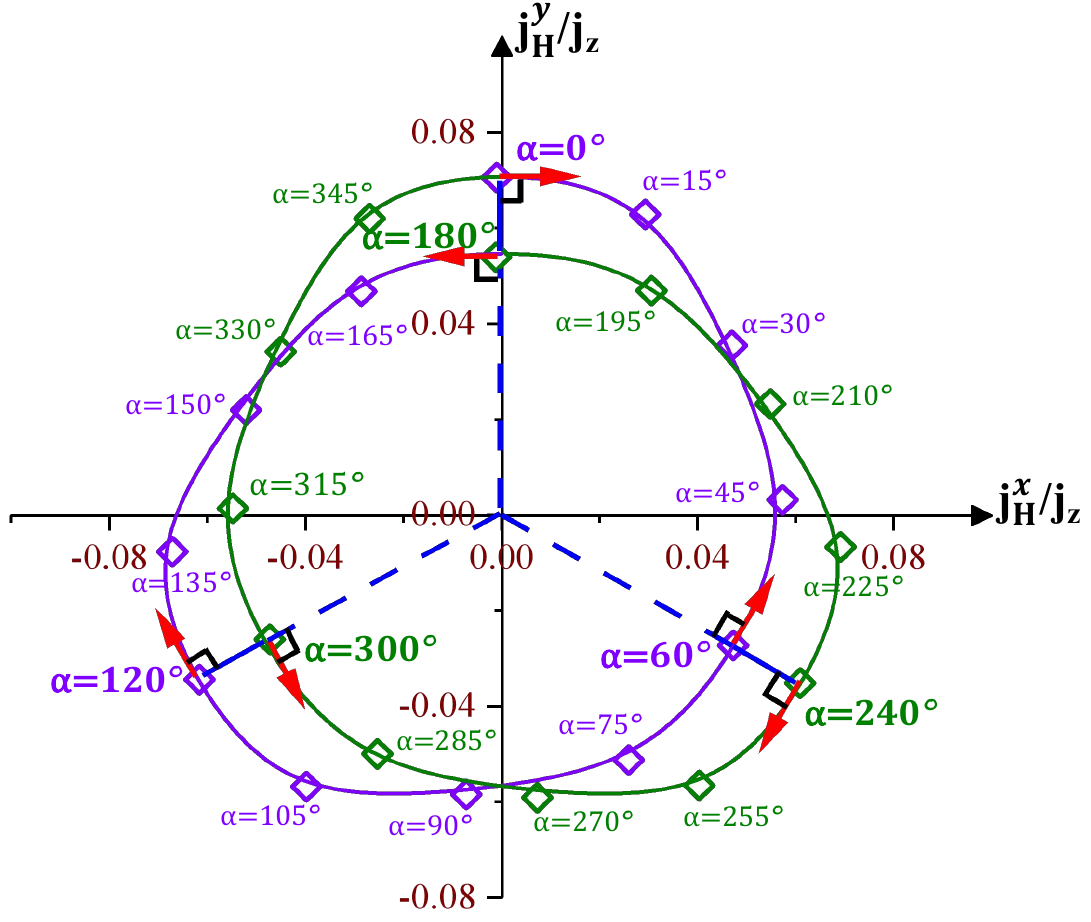}
	\caption{The angle ($\alpha$) dependence of the anomalous Hall currents ($j_H^y$ and $j_H^x$) for the [Cu$_2$$\vert$Co$_2$]$_n$ superlattice along the fcc (111) direction, with all other parameters identical to those in Fig.~\ref{angle}. The rhombuses represent the calculated results, while the solid lines correspond to the fitting results obtained using Eq.~(\ref{fit}) with $n=4$.}
	\label{sc}
\end{figure} 

Moreover, the deviation and chirality of the anomalous Hall current can be further enhanced by constructing a superlattice along the fcc (111) direction, such as [Cu$_2$$\vert$Co$_2$]$_n$, where ``2'' denotes the number of atomic layers and $n$ represents the number of repeating units. As seen in Tab.~\ref{pars}, the contribution from $p_2\simeq0.061$ dominates rather than the conventional AHE from $p_1\simeq-0.0077$, resulting in a situation where the direction of the anomalous Hall current remains unchanged even when the magnetization direction is reversed, as shown in Fig.~\ref{sc}. This demonstrates the manipulation of different components in Eq.~(\ref{fit}) by constructing superlattices and highlights the need for further research into the nonlinear AHE.

\begin{figure}[tp]
	\includegraphics[width=0.8\columnwidth]{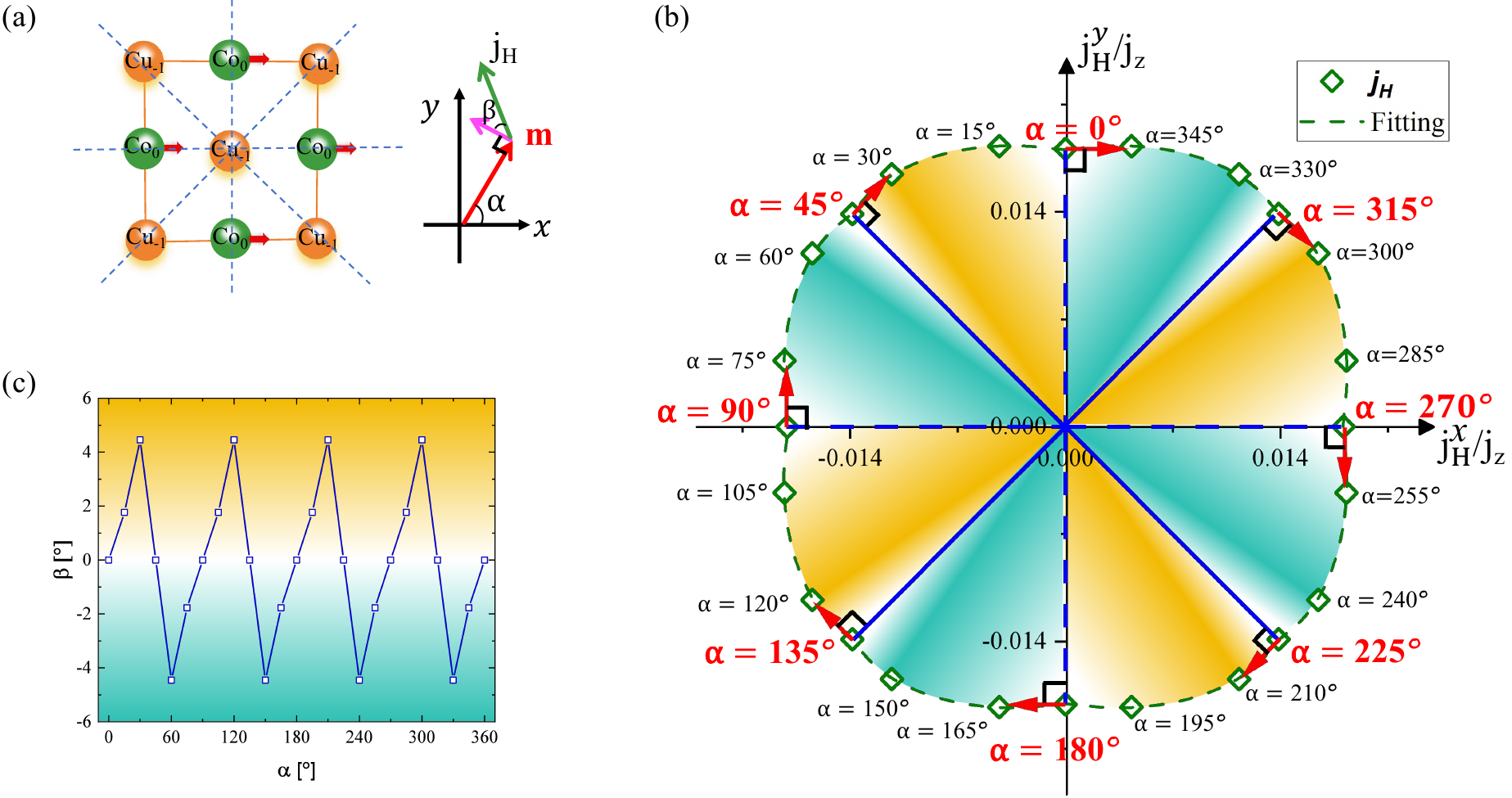}
	\caption{(a) Top view of the Cu$\vert$Co interfaces along (001) direction and the definitions of $\alpha$ and $\beta$, which are the same as those in Fig.~\ref{angle}. (b) The angle ($\alpha$) dependence of the anomalous Hall currents $\mathbf{j}_H^i=(j_H^x,j_H^y)$ for the first Co layer around the Cu$\vert$Co interface. The dashed line represents the fitting results obtained using Eq.~(\ref{fit2}) with $n=3$. (c) The relationship between $\alpha$ and the deviation angle $\beta$.}
	\label{fcc001}
\end{figure}

Based on the previous discussions, the low symmetry of the interface along fcc (111) plane is pivotal for the chirality of AHE. Conversely, the chirality of the AHE is expected to vanish in interfaces characterized by high symmetry, such as the fcc (001) structure. As depicted in Fig.~\ref{fcc001} (a), the in-plane crystal structure of the fcc (001) direction satisfies two mirror symmetries according to $xz$- and $yz$- plane, as sketched by the dash lines, respectively. Moreover, for any $\mathbf{m}$ in the $xy$-plane ($\alpha\in[0,360]$), the projection of $\mathbf{m}$ onto $x$- or $y$- axis ($m_x$ or $m_y$) can be scrutinized using the above mirror symmetry argument along the $xz$- or $yz$- plane, respectively. Consequently, as shown in Fig.~\ref{fcc001} (b), we consistently find $\mathbf{j}_H(\mathbf{m})=-\mathbf{j}_H(-\mathbf{m})$ and the hidden chirality for any $\alpha$. 

However, as illustrated by the solid and dashed blue lines in Fig.~\ref{fcc001} (b), there are two distinct types of primary axes. A crystal-induced transverse current emerges when the Hall current deviates from these primary axes, leading to a misalignment of the corresponding anomalous Hall current ($\mathbf{j}_H\nparallel \mathbf{m}\times\mathbf{j}_c$ with $\beta\neq0$). This phenomenon can also be explained using tensor theory, taking into account the in-plane quadruple rotational symmetry ($C_{4v}$) and the mirror symmetries with respect to $x$, $y$, $y=x$, and $y=-x$ ($\mathcal{M}_x$, $\mathcal{M}_y$, $\mathcal{M}_{y=x}$, and $\mathcal{M}_{y=-x}$). We arrive at the expression,
\begin{eqnarray}
	\mathbf{j}_H/j_z=\left[\begin{array}{c}
		j_H^x/j_z \\
		j_H^y/j_z
	\end{array}\right]=\sum_{i=4n-3}^{n\in[1,\infty)}p_{i}\left[\begin{array}{c}
		\sin i\alpha \\
		-\cos i\alpha 
	\end{array}\right]+\sum_{j=4n-1}^{n\in[1,\infty)}p_{j}\left[\begin{array}{c}
		\sin j\alpha \\
		\cos j\alpha 
	\end{array}\right].
	\label{fit2}
\end{eqnarray}
As shown by the dashed line in Fig.~\ref{fcc001} (b), using $n=3$ is sufficient for an accurate fitting of Eq.~(\ref{fit2}) and the fitting parameters are listed in Tab.~\ref{pars}. Considering the perpendicular condition $\mathbf{j}_H\cdot\mathbf{m}\propto\sin4n\alpha_p=0$, we find that $\alpha_p=l\pi/4$ for integer values of $l$, which is consistent with the results shown in Fig.~\ref{fcc001} (b).

\section*{Conclusion}
In conclusion, we study the AHE symmetrically by considering the discrete symmetry of the solid, rather than relying on the continuum model. Surprisingly, as a consequence of the discrete in-plane rotational symmetry, the direction of the anomalous Hall current is not always follow the rule of $\mathbf{j}_H=\Theta \mathbf{m}\times\mathbf{j}_c$, which highlights the importance of the discrete atomic position in solids for spin transport and thus all related measurements using AHE should be revisited. We also predict the CAHE with IC and demonstrate the general existence of the chirality in spin transport through the interfaces. Due to the different symmetry between the longitudinal and transverse transport, the chirality in the longitudinal conductance is always hidden and can be revealed by the transverse AHE measurements for interfaces with low in-plane symmetry. 


\section*{Supplementary information}

The details, additional references, and extended data are available in the supplementary information.

\begin{acknowledgments}
This work is financially supported by the National Key Research and Development Program of China (Grant Nos. 2023YFA1406600, 2021YFA1202200, and 2020YFA0309600), the University Development Fund of the Chinese University of Hong Kong, Shenzhen, and the Hong Kong Research Grants Council Grants (16302321, 16300522, and 16300523).
\end{acknowledgments}

\bibliographystyle{apsrev4-2}
\bibliography{my.bib}

\end{document}